\documentclass{article}
\usepackage{spconf,amsmath,graphicx}
\usepackage[hidelinks]{hyperref} % URLs
\usepackage{booktabs}   % \toprule, \midrule, \cmidrule, \bottomrule
\usepackage{multirow}   % \multirow
\usepackage{graphicx}   % \resizebox
\usepackage{tabularx}
\usepackage{float}
\usepackage{placeins} % This provides \FloatBarrier, which can be used to make images appear where you want, or at least not too far away.
\usepackage{subcaption}

\usepackage{titlesec}
\titlespacing*{\paragraph}{0pt}{0.25\baselineskip}{0.5em}

\usepackage{tikz}
\usepackage{everypage}
\newcommand\BackgroundText{%
  \begin{tikzpicture}[remember picture,overlay]
    \node [rotate=0, scale=1.3, text opacity=1.0, color=gray] at (current page.south west) [anchor=south west, xshift=0.7cm, yshift=0.9cm] {accepted at ICASSP 2026};
  \end{tikzpicture}
}
\AddEverypageHook{\BackgroundText}
\usepackage[table]{xcolor}
\definecolor{EnCcol}{HTML}{b43b87}   % pink
\definecolor{Mimicol}{HTML}{7f12aa}  % purple
\definecolor{DACcol}{HTML}{376f95}   % blue
\definecolor{XC2col}{HTML}{229b56}   % green

\title{How to Label Resynthesized Audio: \\ The Dual Role of Neural Audio Codecs in Audio Deepfake Detection}
\name{Yixuan Xiao$^{\star}$ \qquad Florian Lux$^{\dagger}$ \qquad Alejandro Pérez-González-de-Martos$^{\dagger}$ \qquad Ngoc Thang Vu$^{\star}$}
  
  \address{$^{\star}$ University of Stuttgart, Institute for Natural Language Processing, Germany \\
      $^{\dagger}$AppTek GmbH, Germany}

\begin{document}
\ninept
\maketitle
\begin{abstract}
Since Text-to-Speech systems typically don't produce waveforms directly, recent spoof detection studies use resynthesized waveforms from vocoders and neural audio codecs to simulate an attacker. Unlike vocoders, which are specifically designed for speech synthesis, neural audio codecs were originally developed for compressing audio for storage and transmission. However, their ability to discretize speech also sparked interest in language-modeling-based speech synthesis. Owing to this dual functionality, codec resynthesized data may be labeled as either bonafide or spoof. So far, very little research has addressed this issue. In this study, we present a challenging extension of the ASVspoof~5 dataset constructed for this purpose. We examine how different labeling choices affect detection performance and provide insights into labeling strategies.
% The abstract should contain about 100 to 150
\end{abstract}
\section{Introduction}
\label{sec:intro}

A Neural Audio Codec (NAC) is originally proposed as a data-driven compression model that maps a waveform to a sequence of compact latent codes and reconstructs the signal from those codes with a learned decoder. 
Similar to classical codecs (e.g., Opus~\cite{valin2016high}, MP3) that are widely used to reduce bitrate and stabilize streaming under bandwidth constraints, NACs also enable efficient transmission. 
More specifically, modern NACs include a vector-quantized encoder-decoder architecture with one or more codebooks. 
% Alex: Maybe cite some pioneering works here already? SoundStream, EnCodec, DAC?
The encoder projects the input audio into frame-level latent representations, which are then quantized by selecting indices from a learned codebook; the \emph{decoder} reconstructs the waveform from these discrete indices, also known as \emph{discrete tokens}. Compared to classical codecs, the discrete tokens from NACs can capture richer information including phonetics content, prosody, and timbre, while removing fine-grained redundancies. 

\begin{figure}[t]  
    \centering
    \includegraphics[width=\columnwidth,clip,trim=8px 8px 8px 60px]{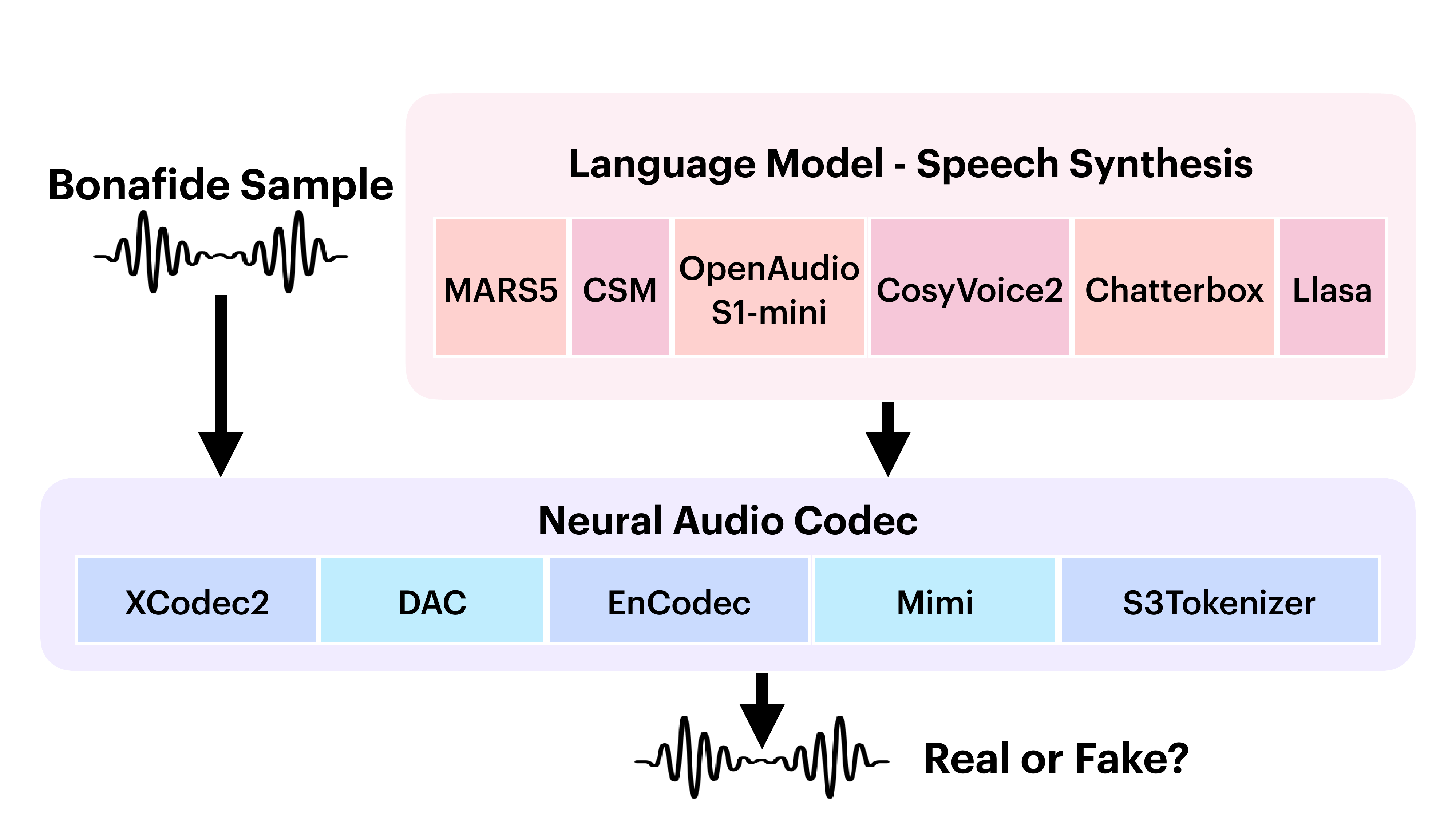}
    \caption{Overview of the setting we investigate: Modern attackers utilize NACs to obtain discrete tokens for language modeling. Human speech is also encoded with NACs for storage or transmission. Real and fake speech sometimes share the same NAC decoders. Can we still detect the fakes?}
    \label{fig:overview}
\end{figure}
% Now from discrete tokens, we move to codec-based speech generation, and the advantage of --> so we can say this will be the trend, and the future
% then we can talk about the challenges, and lead to our motivation
As a result, discrete tokens provided by NACs can be used in many other tasks, such as speech synthesis~\cite{ye2025llasa, baas2025mars6, csm, fish_speech, du2024cosyvoice, du2024cosyvoice2, du2025cosyvoice}. These speech generation systems model the sequence of codec tokens and reconstruct them to waveforms using NAC decoders.
Compared to non-codec-based speech generation methods, codec-based ones often yield higher perceptual quality and speaker similarity. 

Given the advantages of NACs and codec-based speech generation, it is foreseeable that there will be more applications in the future. Hence modern audio deepfake detection systems must handle a spectrum that spans (i) Bonafide speech that has been \emph{codec re-synthesized} (CoRS) by an NAC used purely for transmission, and (ii) \emph{codec-based speech generation} (CoSG) in which a codec-based speech synthesis model produces spoofed audio. 

This dual use introduces a key challenge: how should CoRS be incorporated during training? A naive approach is to treat CoRS either as bonafide or as spoof, given that some NACs are designed for transmission (e.g., EnCodec~\cite{defossez2022high}) and others are more tailored towards speech generation (e.g., Mimi~\cite{defossez2024moshi}).
However, if CoRS is treated as bonafide, the detector may  correctly detect bonafide CoRS but may fail to detect CoSG that uses the same NAC. On the other hand, if CoRS is treated as spoof, the detector may learn to flag codec artifacts as spoof; this can improve accuracy at detecting CoSG~\cite{wu24p_interspeech, chen2025codecfakelargescaleneuralaudio}, but risk rejecting bonafide audio compressed by NACs. 

%In short, the detector only learns to detect artifacts that are mainly codec-specific, hence detecting artifacts is not detecting fakes. We believe that codec-specific artifacts are relatively shallow, signl-level cues. Therefore, a robust detector should emphasize higher-level, codec-invariant information. 

However, only very few studies have examined the dual role of CoRS in audio deepfake detection. CodecFake~\cite{wu24p_interspeech} and CodecFake+~\cite{chen2025codecfakelargescaleneuralaudio} recognized the threat posed by CoSG systems. \cite{chen2025codecfakelargescaleneuralaudio} is also the first to coin the terms CoSG and CoRS. However, they primarily leveraged CoRS as spoof augmentation. Another work on ALM-based deepfakes~\cite{Xie24ALMADD} has analyzed the detectability of some CoSG systems but did not consider CoRS. One main reason of the lack of research is that we currently lack related datasets.

Our contributions include: (i) Releasing an open-source and highly challenging dataset for audio deepfake detection research.\footnote{\url{https://huggingface.co/datasets/Flux9665/CodecDeepfakeDetection}}\footnote{ \url{https://zenodo.org/records/17225924}} This dataset is based on the ASVspoof~5 contribution protocol and can be viewed as an extension of ASVspoof~5~\cite{wang2024asvspoof, wang2025asvspoof}. 
(ii) Investigating the dual role of CoRS to address the research gap, through which we attempt to identify the best labeling strategy for CoSG. Specifically, when to label as bonafide and when to label as spoof. %3) Proposing the use of Barlow Twins to learn codec-invariant artifacts.

\section{Methods}
\subsection{TTS System Selection}

Table~\ref{tab:codeclist} summarizes the selection of attacker systems and NACs considered in this work. All systems were chosen due to their popular open-source implementations and general availability to the public:

\paragraph*{Llasa 8B} Llasa~\cite{ye2025llasa} is a set of LLM-based auto-regressive (AR) TTS models built on top of LLaMa~\cite{grattafiori2024llama}. They extend LLaMa by incorporating speech tokens from XCodec2~\cite{ye2025llasa}. Due to its high parameter count, Llasa~8B offers excellent semantic modeling capabilities. The model generates speech with highly convincing prosodic patterns that align with a message's content.

\paragraph*{MARS5} The MARS series of models~\cite{baas2025mars6} follows a two-stage approach similar to VALL-E~\cite{chen2025neural}. It offers two types of voice cloning: \textit{shallow-clone}, where a speaker embedding is computed from a reference audio and used to condition the speech generation; and \textit{deep-clone}, where in addition to that, the reference audio and its corresponding transcript are included as a prefix of the target sequence. The voice consistency makes MARS5 attractive to attackers. 

\paragraph*{CSM} Sesame's CSM~\cite{csm} is a speech generation model based on LLaMa that produces Mimi~\cite{defossez2024moshi} speech tokens given text inputs. It follows a dual-decoder architecture similar to Moshi~\cite{defossez2024moshi}. %, where the backbone (larger) decoder (\textit{Temporal Transformer}) predicts the $L0$ codebook tokens, and a smaller decoder (\textit{Depth Transformer}) predicts the remaining codebook tokens in an auto-regressive manner. 
The combination of a dual-decoder setup and interleaved text-audio tokens enables low-latency streaming speech generation, making it suitable for real-time conversational attack scenarios. 

\paragraph*{OpenAudio S1-mini} OpenAudio's S1-mini is the successor of the popular Fish-Speech model~\cite{fish_speech}. It supports 13 different languages and offers controllability over a number of emotions, tones and non-verbal cues (such as laughter, sobbing, groaning, etc). This level of controllability makes it attractive for social media or media and entertainment deepfakes.

\paragraph*{Chatterbox} Chatterbox~\cite{chatterboxtts2025} is Resemble AI's open-source multilingual LlaMa-based TTS system supporting 23 languages. 
% It incorporates an audio watermarking method, which is enabled by default. However, experienced users can easily disable this in the inference code.
It excels at reproducing speaker timbre and style from a short reference audio, which does not even need a transcription, unlike all other systems we consider. Together with the simple interface through a Python package, this makes it attractive to non-expert attackers.

\paragraph*{CosyVoice~2} CosyVoice~2~\cite{du2024cosyvoice, du2024cosyvoice2, du2025cosyvoice} is an LLM-based TTS system built on top of \textit{Qwen2.5-0.5B}~\cite{yang2025qwen2}. While the generation happens in two stages and only the first stage works autoregressively, the second stage can be run in a chunked manner, making low latency possible while offering great modularity.

\begin{table}[b]
\centering
\begin{small}
\begin{tabular}{l l l r c}
\toprule
\textbf{Abbrev.} & \textbf{Attacker} & \textbf{Codec} & \textbf{Params.} & \textbf{Bonafide} \\
\midrule
LSA & Llasa 8B     & XCodec2                 & 8B   & Yes \\
MS5 & MARS5        & EnCodec                 & 1B   & Yes \\
CSM & CSM          & Mimi                    & 1B   & Yes \\
OS1 & OA S1-mini & DAC    & 0.5B & Yes \\
CBX & Chatterbox   & S3Tokenizer             & 0.5B & No  \\
CV2 & CosyVoice2   & S3Tokenizer             & 0.5B & No  \\
\bottomrule
\end{tabular}
\end{small}
\caption{\label{tab:codeclist} List of selected TTS and associated NAC models. The Bonafide column refers to whether the codec is also used to compress and decompress bonafide audios.}
\end{table}

\subsection{Neural Audio Codec Selection}
All NACs are extracted from their corresponding TTS systems.
\paragraph*{XCodec2} XCodec2~\cite{ye2025llasa} uses only a single codebook, making it a natural choice for extending pretrained text-only LLMs to the speech generation task. Similar to other recent works, it employs Finite Scalar Quantization (FSQ) to maximize codebook utilization rates.

\paragraph*{EnCodec} Originally proposed for general audio quantization (speech, music, etc.), EnCodec~\cite{defossez2022high} employs Residual Vector Quantization (RVQ) to discretize audio into multiple hierarchical codebooks. Each quantization layer models the residual error of the previous one. This dependency between quantization layers increases the modeling complexity for other components in the pipeline.

\paragraph*{Mimi} Similar to EnCodec, Kyutai's Mimi~\cite{defossez2024moshi} also follows an RVQ quantization method. However, Mimi works at a significantly lower token rate (12.5Hz), and uses a distillation loss to enhance the encoding of semantic information into its first codebook layer. Its low token rate helps reduce computational requirements of next-token prediction models. 

\paragraph*{DAC} Introduced as a refinement over EnCodec, the Descript Audio Codec (DAC)~\cite{kumar2023high} improves audio fidelity and compression rates, keeping EnCodec's RVQ structure while enhancing different aspects of the architecture to achieve overall better performance.

\paragraph*{S3Tokenizer} Unlike most NACs, the S3Tokenizer~\cite{an2024funaudiollm} is an encoder-only speech tokenizer model optimized for phonetic and prosodic content encoding. It is built on top of an Automatic Speech Recognition acoustic model, adding a FSQ layer to its encoder outputs. Thus, to convert speech tokens back into audio, one needs a separate model such as the Flow Matching decoder proposed in CosyVoice~2~\cite{du2024cosyvoice}. The lack of a standardized decoder for this codec rules out its use for pure audio compression and reconstruction, hence we do not apply it to the bonafide samples in our dataset. % Furthermore, both Chatterbox and CosyVoice2 predict spectrograms from the speech encoded with this codec, and then apply spectrogram inversion, which is already many times covered in the base version of ASVspoof~5. 

% \subsection{Barlow Twins}
% \begin{figure}[t]  
%     \centering
%     \includegraphics[width=\columnwidth]{imgs/BT.pdf}
%     \caption{Overview of the Barlow Twins method: We minimize the distance between the unit matrix and the cross-correlation matrix of a decoder latent vector derived from a raw bonafide audio and one derived from the same audio after applying a neural audio codec.}
%     \label{fig:bt}
% \end{figure}
% 
% To mitigate the chances of a detector reacting overly sensitive to bonafide speech that was encoded and decoded with a NAC for storage or transmission, we propose using a method derived from the Barlow Twins approach for self-supervised learning without contrasting pairs~\cite{zbontar2021barlow}. The goal of this approach is to make an encoder (in our case, the detector model) invariant to some non-semantic change in its input (in our case the application of a NAC). We provide a schematic overview in Figure~\ref{fig:bt}. The intuition behind this is that if the main-diagonal contains only ones, the two vectors contain the same information. And if all other values are zeroes, the dimensions of the vectors contain no redundant information. We found in prior works that this approach is well suited as an auxiliary learning objective~\cite{lux2023combining,lux2023controllable}.

%\section{Dataset Construction}
\begin{table}[t]
\centering
\setlength{\tabcolsep}{4pt}
\resizebox{\columnwidth}{!}{%
\begin{tabular}{lccc}
\toprule
 & \textbf{Train} & \textbf{Dev} & \textbf{Test} \\ \midrule
R / F / A & 3,600/14,400/10,800 & 1,200/4,680 /3,600 & 1,200/7,193 /3,600 \\ \midrule
Speakers & 360 & 719 & 907 \\ \midrule
NACs & EnCodec,Mimi,DAC & EnCodec,Mimi,DAC & All \\ \midrule
\# per TTS & 
\begin{tabular}[c]{@{}c@{}}MS5: 3,600 \\ CSM: 3,600 \\ OS1: 3,600 \\ CV2: 3,600\end{tabular} &
\begin{tabular}[c]{@{}c@{}}MS5: 1,170 \\ CSM: 1,170 \\ OS1: 1,170 \\ CV2: 1,170\end{tabular} &
\begin{tabular}[c]{@{}c@{}} MS5: 1,200 \\ CSM: 1,200 \\ OS1: 1,200 \\ CB2: 1,200 \\ CBX: 1,200 \\ LSA: 1,193\end{tabular} \\
\bottomrule
\end{tabular}
}
\caption{Dataset statistics for train, development, and test partitions. R: Real, F: Fake, A: Augmentation, abbreviations see Table~\ref{tab:codeclist}}
\label{tab:dataset_stats}
\end{table}

\begin{table*}[ht]
\centering
\setlength{\tabcolsep}{3pt}
\renewcommand{\arraystretch}{1.1} 

\resizebox{\textwidth}{!}{%
%\begin{tabular}{ll*{13}{c}}
\begin{tabular}{ll
     >{\columncolor{EnCcol!20}}c
     >{\columncolor{Mimicol!20}}c
    >{\columncolor{DACcol!20}}c
    c
    >{\columncolor{XC2col!20}}c
     c c
    >{\columncolor{EnCcol!20}}c
    >{\columncolor{Mimicol!20}}c
    >{\columncolor{DACcol!20}}c
    >{\columncolor{XC2col!20}}c
    c c}
\toprule
\multirow{2}{*}{Model} & \multirow{2}{*}{NAC} & \multicolumn{13}{c}{Evaluation set} \\
\cmidrule(lr){3-15}
& & MS5 & CSM & OS1 & CV2 & LSA & CBX & T-CoSG & EnC & Mimi & DAC & XC2 & T-CoRS & All \\
\midrule
\multicolumn{15}{c}{\textbf{Base Models trained without NAC augmentation}} \\
\midrule
X-AASIST & -- & 2.75\% & 7.50\% & 5.50\%  & 7.83\% & 23.75\% & 9.25\% & 10.67\% & 24.50\% & 31.03\% & 17.75\% & 10.92\% & 22.06\% & 20.35\%  \\
LWBN        & -- &  1.50\% & 4.25\% & 3.42\% & 5.50\% & 24.08\% & 7.33\% & 10.00\% & 20.71\% & 29.75\% & 16.00\% & 9.29\% & 20.65\% & 19.35\%  \\
\addlinespace[4pt]
\midrule
\multicolumn{15}{c}{\textbf{CoRS treated as \emph{bonafide} during training}} \\
\midrule
\multirow{4}{*}{X-AASIST} & \textcolor{EnCcol}{EnC} & -0.08 \% & +4.67\% & -0.83\% & +1.33\% & -5.58\% & +2.75\% & +0.30\%  & -17.92\% & -6.51\% & -4.43\%  & -0.17\% & -7.10\% & -6.12\%  \\  %w2va.enc.acodec
& \textcolor{Mimicol}{Mimi} & -2.17\% & +6.75\% & -3.17\% & -2.00\% & -10.58\% & -1.00\% & -2.00\% & -13.75\% & -17.86\% & -7.67\% & -0.08\% & -10.77\% & -9.56\% \\
& \textcolor{DACcol}{DAC} & -0.75\% & -0.58\% & -2.42\% & -3.17\% & -8.41\% & -0.58\% & -2.81\% & -14.08\% & -9.28\% & -9.21\% & -0.67\% & -8.69\% & -7.98\%  \\
& \textcolor{XC2col}{XC2} & -1.25\% & +0.17\% & -0.50\% & -0.50\% & +7.18\% & +0.25\% & +2.01\% & -2.50\% & -0.86\% & +1.42\% & -0.96\% & -0.33\% & -0.15\% \\
\midrule
\multirow{4}{*}{LWBN} & \textcolor{EnCcol}{EnC} & -0.42\% & +2.50\% & -0.92\% & +0.25\% & -9.17\% & +2.75\% & -1.77\% & -15.97\% & -5.87\% & -6.25\% & -0.62\% & -7.23\% & -6.87\%  \\
& \textcolor{Mimicol}{Mimi} &  -1.00\% & +12.08\% & -2.17\% & -2.42\% & -14.61\% & -1.25\% & -2.46\% & -12.13\% & -18.83\% & -8.41\% & +0.55\% & -11.33\% & -10.40\% \\
& \textcolor{DACcol}{DAC} &  -0.83\% & +1.58\% & -1.33\% & -1.58\% & -10.58\% & +1.75\% & -3.13\% & -12.36\% & -9.17\% & -9.41\% & -0.87\% & -8.85\% & -8.50\% \\
& \textcolor{XC2col}{XC2} &  +0.25\% & +1.58\% & +1.67\% & +0.67\% & +7.60\% & +1.17\% & +1.86\% & +3.12\% & +2.85\% & +3.00\% & -0.12\% & +2.33\% & +2.18\%  \\
\addlinespace[4pt]
\midrule
\multicolumn{15}{c}{\textbf{CoRS treated as \emph{spoof} during training}} \\
\midrule
\multirow{4}{*}{X-AASIST} & \textcolor{EnCcol}{EnC} & -1.08\% & +0.50\% & -0.42\% & +0.67\% & +2.82\% & +1.00\% & +1.76\% & +12.25\% & +1.13\% & +3.67\% & +1.50\% & +5.37\% & +4.73\%  \\
& \textcolor{Mimicol}{Mimi} & -0.00\% & -1.33\% & +1.58\% & +0.17\% & +2.82\% & -0.08\% & +0.83\% & +6.74\% & +5.49\% & +2.76\% & +1.36\% & +4.91\% & +4.37\%  \\
& \textcolor{DACcol}{DAC} & +0.42\% & +4.33\% & +4.83\% & +7.58\% & +7.85\% & +5.17\% & +5.31\% & +10.33\% & +8.08\% & +10.46\% & +3.92\% & +8.54\% & +8.12\%   \\
& \textcolor{XC2col}{XC2} & -0.17\% & +1.67\% & +0.25\% & +1.92\% & -3.25\% & +2.33\% & +0.58\% & -1.42\% & -2.27\% & -0.25\% & +3.42\% & -0.73\% & -0.55\% \\
\midrule
\multirow{4}{*}{LWBN} & \textcolor{EnCcol}{EnC} &  +1.00\% & +0.92\% & +1.67\% & +1.17\% & +6.18\% & +0.75\% & +2.00\% & +13.45\% & +3.81\% & +4.92\% & +2.21\% & +6.44\% & +5.81\%  \\
& \textcolor{Mimicol}{Mimi} & +0.83\% & +0.42\% & +2.00\% & +1.75\% & +6.01\% & +1.92\% & +1.75\% & +5.54\% & +6.44\% & +3.44\% & +1.55\% & +4.81\% & +4.15\% \\
& \textcolor{DACcol}{DAC} & +1.17\% & +1.83\% & +4.08\% & +3.75\% & +9.00\% & +2.17\% & +3.12\% & +9.70\% & +6.19\% & +7.86\% & +2.05\% & +7.03\% & +6.45\% \\
& \textcolor{XC2col}{XC2} &  -0.08\% & +0.50\% & -0.00\% & +0.50\% & -3.46\% & +0.42\% & -0.83\% & +0.79\% & -0.50\% & -0.39\% & +0.40\% & -0.08\% & -0.37\% \\
\bottomrule
\end{tabular}%
}
\caption{Impact of labeling CoRS as bonafide vs.\ spoof during training. The color of each column indicates whether the result is associated with a particular NAC. e.g., the MARS5 (MS5) column is colored the same as \textcolor{EnCcol}{EnCodec}, since it uses its decoder for speech generation.}
%The rows can be divided into three groups: base models trained without augmentation, models treating CoRS as bonafide augmentation, and models treating CoRS as spoof augmentation. 
\label{tab:dual_role}
\end{table*}

\subsection{Dataset Statistics}

%The statistics of the constructed dataset can be seen at Table \ref{tab:dataset_stats}. 
Table \ref{tab:dataset_stats} shows the statistics of the CodecDeepfakeDetection (CDD) dataset. 
To align with the data generation process in \textit{ASVspoof~5}, we ensured that the distribution of bonafide and spoof speakers follow the same overlap rule, i.e., only half of the speakers overlap between the raw bonafide and CoSG spoof partitions. 

For each NAC, we resynthesize 3.6k/1.2k/1.2k samples for train/dev/test. 
During training, samples are resynthesized from the same bonafide set to meet augmentation needs. 
For dev/test, we instead use 1.2k unique bonafide utterances per NAC to avoid bias from repeated resynthesis.

%Similarly, 
%For each NAC,  we reconstruct 3.6k/1.2k/1.2k samples for train/dev/test, respectively.  
%For the train set, due to augmentation requirements, i.e.,  replacing bonafide samples with a certain probability,
%, or ensuring in the Barlow Twins algorithm that each bonafide sample has a paired reconstructed version—
%all NACs reconstructed from the same bonafide set. 
%In other words, we obtained four reconstructed versions sharing identical content.
%For dev/test sets, to avoid bias caused by repeated reconstructions of identical content, we randomly sampled 1.2k bonafide utterances for each NAC and reconstructed them individually.

%\subsection{Detection Models}

%\subsection{Barlow Twin Augmentation}
% Do we have space?

\section{Experiments}

\subsection{Model Selection and Training}

We used XLS-R\footnote{https://huggingface.co/facebook/wav2vec2-xls-r-300m} as frontend with two backends: AASIST~\cite{jung2022aasist} and LWBN~\cite{xiao25d_interspeech}, later referred to X-AASIST and LWBN as our detectors. Both have shown strong performance on the ASVspoof Challenge or In-the-Wild~\cite{muller22_interspeech} datasets.
All audio samples were adjusted to 4 seconds. For samples shorter than 4 seconds, looping was applied before a random 4-second segment was extracted. Augmentation was applied with probability 0.3 for RIRNoise\cite{ko2017study} and 0.2 for RawBoost~\cite{tak2022rawboost}. Training was carried out for max\_epoch 50 with a learning rate 1e-4, using a StepLR scheduler (step\_size=20, $\gamma$=0.5). Early stopping was applied with patience=10, min\_$\delta$=1e-3. Training scripts and codebase are released.\footnote{https://github.com/XIAOYixuan/IMS-ADD/tree/codec-add}

\subsection{Direct Inference}

%In this experiment, we assess the difficulty of CDD. We first train two detectors on the ASVspoof~5 training set, and then test directly on the CDD dataset. Next, we train on CDD training set, which only contains \emph{raw bonafide} audio (without re-synthesized speech), and then test on CDD.
%We define three types of test sets: (1)\textbf{T-CoSG}: Raw bonafide and TTS-generated data. (2) \textbf{T-CoRS}: Re-synthesized bonafide and TTS-generated data. (3) \textbf{Both}: All data, including raw bonafide, re-synthesized bonafde, and TTS-generated data.  % We also evaluate models on ASVspoof~5's test st (track 1).

In this experiment, we assess the difficulty of CDD. Two detectors are first trained on ASVspoof~5 and tested on CDD. Then, two new detectors are trained on CDD without CoRS augmentation.
We evaluate on three types of test sets: (1) \textbf{T-CoSG}: raw bonafide + CoSG spoof; (2) \textbf{T-CoRS}: CoRS bonafide + CoSG spoof; (3) \textbf{All}: all data (raw bonafide, CoRS bonafide, CoSG spoof).

\subsection{Investigating the Dual Roles of CoRS}

In this experiment, we study how treating CoRS as either bonafide (NAC for transmission) or spoof (NAC for speech generation) during training affects detection performance.
Thus, for each detector, we train four variants. Each variant uses a single NAC for augmentation: with probability $0.5$, a raw bonafide sample is replaced with its resynthesized version. 
For example, with Mimi, each raw bonafide has a 50\% chance of being resynthesized by Mimi. If CoRS is treated as bonafide, the label remains unchanged during training; otherwise, the label is switched to spoof. Note that the label changes occur only during training; dev and test labels remain fixed as bonafide.

Results are reported on T-CoSG, T-CoRS, and All. In addition, T-CoSG and T-CoRS are further decomposed into more fine-grained subsets to understand the insights. Specifically, for T-CoSG, we report the EER on all raw bonafide plus spoof data generated by a single TTS (e.g., all raw bonafide + MS5 spoof). For T-CoRS, we report EER on bonafide resynthesized by a single NAC plus all spoof data (e.g., Mimi bonafide + spoof data from all TTS systems).

%We hypothesize that when NAC-generated CoRS is treated as spoof, its corresponding test performance will degrade due to label conflicts. However, we are also interested about the performance on other T-CoRS subsets, such as recognizing bonafide resynthesized by EnCodec.

%\subsection{Augmentation Experiments}

\section{Results and Discussion}

\subsection{A New Challenge: Codec-based Attacks }

Table \ref{tab:direct_infer} shows the Equal Error Rate (EER). It can be observed that our proposed dataset has a very different data distribution compared to ASVspoof~5. 
%Although both detectors achieve good performance on ASVspoof~5, their performance on CDD is significantly worse. 
Knowledge learned on one dataset %, and the performance achieved,
does not transfer well to another.
After training two detectors with CDD's training set, their performance on T-CoSG improves notably; however, the EER remains relatively high on the \textit{All} set, reaching 20.35\% and 19.35\%, respectively. Moreover, when we replace the bonafide sets of T-CoSG with resynthesized speech, the performance on T-CoRS degrades across all detectors, with EER increasing by at least 9.79\%. Hence, codec-based attacks pose a novel challenge. 

Table \ref{tab:dual_role} shows more detailed results for models trained on CDD but without NAC augmentation. Among all TTS systems, MS5, CSM, OS1, and CV2 are seen during training, while LSA and CBX are not. However, the EER is significantly worse for LSA ($>$ 20\%) but not for CBX ($<$ 10\%). We attribute this to the fact that both CV2 and CBX adopt the S3Tokeizer, leading to shared characteristics. As a result, training on one provides better generalization to the other TTS system. 

The performance on T-CoRS is opposite to T-CoSG. On T-CoRS, NACs can be grouped by training exposure:
(i) \emph{partially seen} NACs (EnC, Mimi, DAC), whose corresponding TTS attacks appear in training; (ii) \emph{unseen} XC2.
The EER is notably higher on partially seen NACs than on the unseen. This indicates that the models capture \emph{codec-specific} artifacts. As a result, achieving good performance on codec-based attacks carries \emph{greater risk} of misjudgment if the same NACs are also used during transmission.
% NOTE, experiment IDs
% wav2vecAASIST: assist.asv5
% LWBN: lwbn.asv5
% w2vassist cdd: assist.acodec

\begin{table}[t]
\centering
%\begin{tabular}{l|c|c|c|c}
\begin{tabular}{l c c c c}
\toprule
Model & ASV5 & T-CoSG & T-CoRS & All \\ % two test set, without codec or with
\midrule
%AASIST &  &  &  &  \\ TODO
%RawNet &  &  &  &  \\
X-AASIST &  9.07\% & 33.09\%  & 45.33\%  & 43.13\%  \\ % assist.asv5: acodec-fast 
LWBN & 8.27\% & 31.24\%  & 41.03\%  & 39.29\%   \\ % lwbn.asv5.: acodec-fast
X-AASIST (CDD) & 51.63\% & 10.67\% & 22.06\%  & 20.35\%  \\  % assist.acodec: acodec-fast
LWBN (CDD) & 68.02\%  &10.00\% & 20.65\% & 19.35\%  \\ % lwbn.acode: acodec-fast
\bottomrule 
\end{tabular}
\caption{EER on four test sets. ASV5 refers to ASVspoof~5's test set (track 1). Models (CDD) are trained only on CDD.}
% Alex: Maybe make the caption a bit more informative? Should we mention EER as the metric in the table caption?
\label{tab:direct_infer}
\end{table}

\subsection{CoRS as Bonafide Augmentation}

Results are represented by the difference in EER between the model trained with and without NAC data in Table \ref{tab:dual_role}. As expected, when an NAC is used for bonafide augmentation, its corresponding T-CoRS subset shows reduced EER (e.g., train with EnC bonafide augmentation and test on T-CoRS's EnC subset).
%. e.g., when the model is trained with EnC bonafdie augmentation, the EER on T-CoRS's EnC subset decreased by 17.92\%. 

%Based on the results on T-CoRS dataset, we can further partition NACs into two groups by training exposure: 

For the partially seen NACs, using any one NAC for augmentation greatly lowers EER on the other partially seen NAC subsets  (typically $\geq$ 5\%, sometimes $\geq$ 10\%) for both detectors. The reason might be: without NAC augmentation, all codec-based attacks are labeled as spoof, so codec-specific artifacts become strongly correlated with the spoof label. The model learns \emph{joint}, \emph{codec-specific} cues shared across partially seen NACs, which inflates EER on corresponding subsets. Adding one NAC resynthesized bonafide breaks the correlation: the model sees codec artifacts in both bonafide and spoof, reducing codec-specificity and normalizing performance. 
In contrast, no TTS in the training uses XC2, adding an XC2 bonafide set does not break the artifact-spoof association, and models still perform poorly on the other T-CoRS subsets.

%When augmenting with XC2, we observe a divergence between models: on LWBN, EER increases on the non-XC2 T-CoRS subsets, whereas X-AASIST does not show this degradation. We attribute this to differing objectives. X-AASIST is purely discriminative (binary cross-entropy loss); adding a new bonafide set minimally affects the decision boundary for seen spoofs. By contrast, LWBN adopts one-class learning and learns a set of centers to represent bonafide data. Prior work indicates resynthesized audio lies between bonafide and TTS distributions; thus, adding an XC2 bonafide set likely shifts the learned bonafide centers toward XC2-specific artifacts. The model may then prefer resynthesized audio with XC2 artifacts as bonafdie, pushing non-XC2 resyntheses away from the center and increasing EERs.

For the performance on the T-CoSG subsets, we hypothesized that NAC-based bonafide augmentation would hurt performance on its paired TTS attacks. This holds only for Mimi: adding Mimi bonafide data causes a clear drop on T-CoSG's CSM subset for both detectors. For EnC and DAC, performance slightly improves. This might be due to CSM detection's heavy reliance on codec artifacts: Mimi is designed mainly for TTS; when used for resynthesis, it likely discards information irrelevant to the TTS generation objective,  leaving stronger TTS cues. Labeling these as bonafide confuses the model, making it misclassify some CSM attacks.

%\noindent\textbf{CoRS as Spoof Augmentation}
\subsection{CoRS as Spoof Augmentation}
For the T-CoRS subsets, when a partially seen NAC is labeled as spoof, the detector tends to classify other partially seen NACs as spoof. This aligns with our prior observation: the model has learned joint, codec-specific characteristics, so marking any one as spoof strengthens the artifact-spoof association. In contrast, labeling XC2 as spoof simply adds another attack type; its impact on recognizing other NACs is comparatively minor for both detectors.

For the T-CoSG subsets, we initially expected that marking an NAC as spoof would help detection of TTS using that NAC. This is not consistently true, especially for LWBN, which performs worse on nearly all subsets.
To investigate, we analyzed how the augmented models score bonafide vs.\ spoof samples on T-CoSG. The results are shown in Table \ref{tab:diff_results}, higher scores indicate greater similarity to bonafide. Overall, bonafide scores decrease. LWBN adopts one-class learning and focuses modeling the bonafide class. Introducing more realistic CoRS substantially shifts the learned bonafide centers towards spoof clusters. X-AASIST adopts cross entropy loss. Although the focus is more balanced, the model augmented with DAC shows a significant drop in bonafide scores, meaning many bonafide samples appear ``more fake''. Compared to other NACs, DAC is expressly designed to improve audio fidelity for audio compression; hence it may retain more information from the raw bonafide. This explains why labeling DAC resynthesis as spoof greatly degrades performance for both detectors.

\begin{table}[t]
\centering
\setlength{\tabcolsep}{6pt} 
\begin{tabular}{lcc|cc}
\toprule
 & \multicolumn{2}{c|}{X-AASIST} & \multicolumn{2}{c}{LWBN} \\
\cmidrule(lr){2-3} \cmidrule(lr){4-5}
Class & Spoof & Bonafide & Spoof & Bonafide \\
\midrule
Base &  -4.8688 & 4.5151   & 0.0280   & 0.8702 \\
EnC  &  +0.1765 & -1.1232  & -0.0159  & -0.0495 \\
Mimi &  +0.2001 &  +1.5511 &  +0.0136 & -0.0359 \\
DAC  & -0.4163  & -3.8287  &  +0.0779 & -0.0528 \\
XC2  & +1.8407  & +0.9545  & -0.0013  & -0.0250 \\
\bottomrule
\end{tabular}
\caption{Differences from Base Models for X-AASIST (Average Logit Score) and LWBN (Average Cosine Similarity). }
\label{tab:diff_results}
\end{table}

\section{Conclusion}

% when both exists (NAC-based TTS and NAC resyn)
% should consider treating NAC as bonafide to break the correlation
% make the model codec-artefact invarient
% that can help when the biggest spoofiness doesn't 
% come from codec based artefacts
% but 

We introduce a new and challenging codec-based dataset called CodecDeepfakeDetection. We conduct experiments on this dataset to study the dual role of NACs in augmentation. We find that the design objective of the NAC matters.
For compression-oriented NACs, treating CoRS as spoof risks shifting detectors to misjudge bonafide speech;  % especiall for one-class learning?
when the corresponding CoSG appears in training, using that NAC for bonafide augmentation promotes learning of codec-invariant cues. 
For synthesis-oriented NACs, resynthesis aligns with speech synthesis objectives; labeling such CoRS as bonafide is less effective than labeling them as spoof, which enables the detector to better exploit codec-specific artifacts. The key challenge is posed by NACs that excel at both compression and synthesis; more research will be needed to disentangle codec-artifacts from spoofness.

\vfill
\paragraph*{Acknowledgments:} We would like to thank the ASVspoof committee and in particular Xin Wang for his support and advice.

%Looking forward, the most challenging NACs for audio deepfake detection will be those that are simultaneously strong compressors and highly effetive for speech synthesis. ...
% show what??
% for bonafide, yeah it help promoting code-invariant
% but depends on the design, like Mimi
% if the spoofness is more codec-specific or NACs carries more 
% attack cues (might be cause by the original design)
% , labeling CoRS can harm performance (cuz now 
% fake is more real)
% if spoof, DAC is designed to retain more info, 
% labeling them as fake also harm performance
% cuz real is more fake
% especially bad for one-class learning seem to be more 
% sensitive due to the realism? of the audio
% 
\pagebreak 

\bibliographystyle{IEEEbib}
\bibliography{refs}

\end{document}